\documentclass{aa}
\usepackage{graphicx}
\usepackage{multirow}
\usepackage{bm}
\usepackage{txfonts}
\usepackage{natbib}
\bibpunct{(}{)}{;}{a}{}{,} 

\begin{document}

\title{A method for space-variant deblurring with application to 
adaptive optics imaging in astronomy}

\author{A. La Camera\inst{1}\and
        L. Schreiber\inst{2}\and
        E. Diolaiti\inst{2}\and
        P. Boccacci\inst{1}\and
        M. Bertero\inst{1}\and
        M. Bellazzini\inst{2}\and
        P. Ciliegi\inst{2}
        }

\institute{
Dipartimento di Informatica, Bioingegneria, Robotica e Ingegneria dei Sistemi
(DIBRIS), Universit\`a di Genova, Via Dodecaneso 35, 16145 Genova, Italy\\
\email{[andrea.lacamera|patrizia.boccacci|mario.bertero]@unige.it}
\and
INAF - Osservatorio Astronomico di Bologna, Via Ranzani 1, 40127 Bologna, 
Italy\\
\email{[laura.schreiber|emiliano.diolaiti|michele.bellazzini|paolo.ciliegi]@oabo.inaf.it}
}

\titlerunning{Space-variant deblurring}
\authorrunning{La Camera et al.}

\offprints{Andrea La Camera, \email{andrea.lacamera@unige.it}}

\date{Received --; accepted --}

\abstract  
{Images from adaptive optics systems are generally affected by significant distortions of the point spread function (PSF) across the field of view, depending on the position of natural and artificial guide stars. Image reduction techniques circumventing or mitigating these effects are important tools to take full advantage of the scientific information encoded in AO images.}
{The aim of this paper is to propose a method for the deblurring of the
astronomical image, given a set of samples of the space-variant PSF.}
   {The method is based on a partitioning of the image domain into
regions of isoplanatism and on applying  suitable
deconvolution methods with boundary effects correction to each region.}
   {The effectiveness of the boundary effects correction is proved. Moreover, the criterion for extending
   the disjoint sections to partially overlapping sections is validated. The method is applied to simulated images of a stellar system characterized by a spatially variable PSF. We obtain  good photometric quality, and therefore  good science quality, by performing aperture photometry on the deblurred images. The proposed method is implemented in IDL in the Software Package ``Patch'', which is available on {\texttt{http://www.airyproject.eu}}.}
   {}


\maketitle


\section{Introduction}
\label{intro}

The problem of image deblurring in the case of anisoplanatism of the imaging
system is an important problem in several domains of applied science. In this
paper, we focus on the case of a telescope equipped with an adaptive optics
(AO) system. A basic AO system \citep{beckers1993} includes a deformable mirror, which 
compensates for the time-evolving effects of the atmospheric turbulence and 
other disturbances distorting the optical wavefront of the observed science 
target. The compensation is calculated by a real-time control system on the 
basis of measurements of the disturbances performed on a guide source, for 
instance, a natural star. 

The goal of an AO system with these features, also known as single-conjugate 
adaptive optics (SCAO), is to make the guide star wavefront flat. The science 
target is usually not coincident with the guide star: the light beams from 
the science target and from the guide star cross different volumes of atmosphere 
and therefore are affected by different wavefront aberrations because of the 
stratified structure of the atmospheric turbulence. Therefore even a perfect 
instantaneous correction on the guide star wavefront is not perfect for the 
science target. As a consequence of this mismatch, the point spread function 
(PSF) in the direction of the science target is degraded and typically 
elongated towards the guide star PSF. The PSF elongation across the field of 
view increases with the angular distance from the guide star itself and the 
elongation pattern is approximately radially symmetric with respect to the 
guide star direction \citep{schreiber2012}. 

More complex adaptive optics techniques have been proposed and demonstrated 
to improve PSF uniformity across the field of view. For instance, 
multi-conjugate adaptive optics (MCAO) \citep{beckers1988,marchetti,rigaut} 
is based on the use of multiple deformable mirrors following the stratified 
structure of the atmospheric turbulence and on the use of multiple guide 
stars to reconstruct a kind of three-dimensional mapping of the turbulence 
itself. Despite the remarkable performance uniformity with respect to SCAO 
systems, even in MCAO some residual PSF variation in the field of view (FoV) is 
possible, partly correlated with the position of the guide stars and thus 
following a non-radially symmetric variation pattern.
In summary, depending on the AO flavour and on the necessary degree of PSF 
stability imposed by science requirements, space variation of the PSF in AO 
observations could be a crucial issue to be addressed by image processing methods.

The case of a space-variant PSF, varying from pixel to pixel of the image
domain, is not computationally tractable. However, if the PSF is not too
rapidly varying, it is possible to decompose the domain into patches where
the PSF can be assumed to be approximately space invariant so that
the imaging operator is locally described by a convolution product. 

In this case, it has been proposed to separately deconvolve the different
patches \citep{trussela,trusselb} and to reassemble  the results to 
obtain the final reconstructed image. The difficulty of this approach, 
which we call the sectioning approach, 
is mainly due to boundary artifacts at the boundaries of the different patches
and discontinuities because of the use of different PSFs. To circumvent this
difficulty, the use of partially overlapping patches has been proposed
\citep{boden,aubailly}. However, another approach, which we call the 
interpolation approach, is intended to suppress effects because of the 
discontinuity of the PSF from patch to patch by a suitable interpolation of 
the available samples of space-invariant PSFs. Two different kinds of 
interpolation have been proposed: the first is based on the interpolation of the 
results obtained by convolving the original object with the PSF samples 
\citep{nagy}, and the second is obtained by directly interpolating  the PSF samples 
\citep{hirsch}. Both are considered in \citet{gilad}, while in \citet{denis} 
the authors show that the second kind of interpolation provides more reliable 
results. A similar approach based on the Richardson-Lucy 
(RL; \citet{richardson,lucy}) is proposed in \citet{lauer}. In all approaches, 
fast deblurring is obtained by applying fast Fourier transform (FFT) to space-invariant sub-problems.

We propose an improvement to the sectioning approach. First, we 
introduce a criterion for extending each one of the non-overlapping sections,
corresponding to different PSFs, to a suitable broader section with the same PSF.
Second, we apply  a deconvolution
method with boundary effects correction proposed in 
\citet{bertero2005,anconelli2006} to each one of the new overlapping sections. 
This method is implemented both  in RL 
and in a fast deconvolution method, called scaled gradient projection 
(SGP; \citet{bonettini2009}). All the methods we propose are 
implemented in IDL in a dedicated software package called
``Patch'', described in \citet{ciliegi2014} and freely downloadable from the website 
\texttt{http://www.airyproject.eu}. 

The paper is organized as follows. In Sect.~\ref{method} we describe the adopted 
deconvolution approach referring to the sectioning of the image domain 
(Sect.~\ref{sectioning}). Then, we illustrate the specific 
deconvolution algorithms with boundary effects correction (Sect.~\ref{dec}).
The performance achievable with our approach is illustrated through 
the SGP-based deconvolution of an 
\textit{Hubble} Space Telescope (HST) pre-COSTAR simulated image (Sect.~\ref{hst_example}).  
In Sect.~\ref{aocase} we describe the simulation of a stellar 
AO field both in $J$ and $K_s$ bands. Details on frames generation, 
specifying the science case and the adopted instrument, 
are given in Appendix A, while the PSF model is given in Appendix B.
In Sect.~\ref{reduction} we show how the synthetic images were deconvolved 
(again by means of SGP) and analyzed. 
In Sect.~\ref{results} we report our results: the 
analysis of the reconstructed images (Sect.~\ref{analysis}) and the dependence of the 
photometric and astrometric measurements quality on the spatial knowledge of the PSF 
(Sect.~\ref{accuracy}) also showing  the derived colour-magnitude diagram (CMD). 
The comparison of the CMDs obtained by 
dividing the FoV in an increasing number of sub-domains with 
one obtained by performing a single deconvolution, clearly 
illustrates the improvements in photometric quality offered by our method. 
We summarize our conclusions  in Sect.~\ref{conclusions}.


\section{Method}
\label{method}

The starting point, as in most papers on space-variant deblurring, is to
assume we have $K_{0} \times K_{0}$ samples of the PSF, with centres in points 
${\bf n_1, n_2,..., n_{K_{0}^2}}$ of the image domain. We assume that the
size of the image is $N_0 \times N_0$. The PSF samples can be obtained from a 
model of the space-variant PSF or extracted from the detected image, whenever
this is possible. The problem of PSF extraction and modelling is not trivial and is
beyond the scope of this paper. A brief discussion is reported in Sect.~\ref{conclusions}.

\subsection{Sectioning of the image domain}
\label{sectioning}

For simplicity, we consider the case in which the central points of the PSF
samples form a uniform grid, symmetric with respect to the centre of the
image. If the image size $N_0$ is not divisible by $K_{0}$  then we extend the
image by zero padding to an image $N \times N$ such that $n = N/K_{0}$ is an
integer number. In this way, the image has been sectioned in $K_{0} \times K_{0}$
non-overlapping patches (sections), each one with size $n \times n$. The PSF 
centred in one section is associated with that section and assumed space invariant
across it. 

Besides the problem of boundary effect corrections, which is treated in
the next sub-section, an additional problem is generated by the fact that
different PSFs are associated with adjacent sections; therefore even if we
consider a case of slowly varying PSFs, the deconvolution of disjoint
domains  certainly introduces discontinuities at the common boundaries.
Therefore we extend the disjoint sections to partially overlapping sections. 
The choice of this overlap value is an important parameter and depends 
above all on the extent of the PSFs. To compute it automatically, we define the following positive quantities:
\begin{itemize}
\item $n_P$ is the size of the PSF array;
\item $n_{EE}$ is the size of the sub-array (hence $n_{EE} \leq n_P$) that 
contains the enclosed energy (EE), computed by considering the sum 
of the PSF values inside a squared domain containing 95\% of the total energy;
\item $\Delta n$ is the largest between $n_P -n$ and $n_{EE}$\footnote{In 
order to keep our method (and therefore our software) more general as possible, 
we choose $\Delta n$ even if we know that, very often in AO cases, 
the overlap choice will be driven by $n_{EE}$.}.
\end{itemize} 
Thus, we enlarge each section by taking $n'=n+\Delta n$ as the size of the 
overlapping sections and the resulting total size of the image to be processed 
is therefore $N' = N +\Delta n$. In the software package ``Patch'' a 
larger user-defined $\Delta n$ is also allowed if one must take 
specific features of the image into account.

\subsection{Deconvolution method}
\label{dec}

If we deconvolve the previously defined sections by means of an FFT-based
method, we may obtain boundary artifacts in the form of Gibbs oscillations, 
because, as a consequence of the periodic continuation implicit in the FFT 
algorithm, discontinuities are introduced at the boundaries. Moreover, thanks 
to the PSF extent, the images of stars close to the boundary are not 
completely contained in the image domain or this domain may contain part of 
the images of stars outside the boundary. In the case of iterative methods 
these artifacts can propagate inside the image domain with increasing number 
of iterations so that the reconstruction is completely unreliable.

Since we intend to use an accelerated version of the RL method, we first
consider a simple modification of this method, proposed in \citet{bertero2005},
which compensate, in a simple way, for the boundary effects.

If we denote the section domain as $S$, we then introduce a `reconstruction 
domain' $R$, broader than $S$ and containing all the stars, which, in principle, 
contribute to the image in $S$ as an effect of the PSF extent. 

We assume $K$ is the space-invariant PSF (extended to $R$ by zero padding if 
required) and $A$ is the matrix defined by $Af = K*f$. Moreover, we assume $g$ is 
the image defined on $S$ and extended to $R$ by zero padding; we denote as 
$M_S$ the `mask' of $S$ in $R$, i.e. the function which is 1 on $S$ and 0 
elsewhere. Finally we assume that the image is affected by a background $b$, 
which is assumed to be known. 
Then the modified RL algorithm proposed in \citet{bertero2005} is as follows:
\begin{itemize}
\item define the function
\begin{equation}
\alpha({\bf n}) = \sum_{{\bf n'} \in S}K({\bf n} - {\bf n'}) = 
(A^T M_S)({\bf n})~~,~~{\bf n} \in R~~;
\end{equation}
\item given a threshold $\tau$ set
\begin{equation}
w({\bf n}) = \frac{1}{\alpha({\bf n})}~~{\rm if}~~\alpha({\bf n}) \geq \tau~;
0 ~{\rm elsewhere~ in}~R~~;
\end{equation}
\item given $f^{(0)} > 0$, for k = 0, 1, .... compute
\begin{equation}
\label{RL}
f^{(k+1)} = w \circ f^{(k)} \circ \left ( A^T \frac{ g}{A f^{(k)} + b} \right )~~,
\end{equation}
until a stopping rule is satisfied. 
\end{itemize}
In the previous equations the symbol $\circ$ denotes point-wise product of 
arrays and similarly the quotient symbol denotes point-wise quotient of two
arrays. 

Since we are considering mainly star systems, the algorithm can be 
pushed to convergence \citep{bertero2009}, the limit being a minimizer
of the negative logarithm of the likelihood function for Poisson data, given by
\begin{equation}
\label{KL}
J(f;g) = \sum_{{\bf n'} \in S}\big\{(A f + b)({\bf n'}) - 
g({\bf n'}){\rm ln}[(A f + b)({\bf n'})]\big\}~~. 
\end{equation}
Iterations are stopped when the 
relative variation of this objective function is smaller than a given 
threshold. 

Since the RL algorithm is too slow, a faster convergence is obtained by 
applying the so-called scaled gradient projection (SGP) method 
\citep{bonettini2009}, which is a scaled gradient method as RL since 
the gradient of the objective function (\ref{KL}) is given by
\begin{equation}
\nabla J(f;g) = \alpha - A^T \frac{g}{Af + b}~~. 
\end{equation}
The SGP version including boundary effect correction is given in 
\citet{prato2012} and therefore, for details, we refer to this paper. Here 
we only recall that, if we introduce the following scaling matrix at
iteration $k$ 
\begin{equation}
D_k = {\rm diag}\big({\rm min}
\big[L_2,{\rm max}\big\{L_1,w\circ f^{(k)}\big\}\big]\big)~~,
\end{equation}
where $L_{1}, L_{2}$ are given lower and upper bounds, 
then the descent direction is given by
\begin{equation}
d^{(k)} = P_+ \big(f^{(k)} - \gamma_k D_k \nabla J(f^{(k)};g)\big) - f^{(k)}~~,
\end{equation}
where $P_+$ is the projection on the non-negative orthant and $\gamma_k$ is a
suitable step length, selected according to rules described in 
\citet{prato2012}. Finally, the iteration $f^{(k+1)}$ is obtained by a 
line search, based on Armijo rule, along the descent direction,
\begin{equation}
f^{(k+1)} = f^{(k)} + \lambda_k d^{(k)}~~.
\end{equation}
As shown in \citet{prato2012} this algorithm provides a speed-up 
between 10 and 20 with respect to the RL method. In 
``Patch'' both algorithms are implemented.

The previous algorithms are based on the assumption that data are affected
by Poisson noise, while it is know that they are also affected by read-out
noise (RON), which is described by an additive Gaussian process with zero
mean and variance $\sigma^2$. As shown in \citet{snyder1995} it is possible to
approximate the RON by a Poisson process by adding $\sigma^2$ to both the
data and the background. With this simple modification, the previous 
deconvolution methods can also compensate for the RON effect.

Finally the global reconstructed image is obtained as a mosaic of the 
non-overlapping sub-sections cropped from  reconstructed sub-sections. 
The correctness of photometric and astrometric data in points close to 
the boundaries, as demonstrated by the analysis of our reconstructed 
images, is due to the robustness of RL-like methods with respect 
to (small) errors in the PSF.

\subsection{A test example}
\label{hst_example}

\begin{figure}
\centering
\includegraphics[scale=0.36]{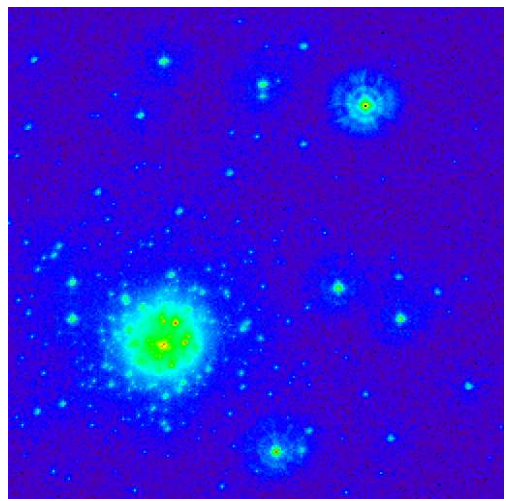}
\caption{Simulation of an observation of a star cluster through HST before 
COSTAR correction.}
\label{fig1}
\end{figure}

As an example of the results achievable with the previous approach, we 
consider the reconstruction of a simulated image of HST before COSTAR 
correction\footnote{obtained via anonymous ftp from \texttt{ftp.stsci.edu} in 
the directory \texttt{/software/tables/testdata/restore/sims/starcluster/}}
(see Fig.~\ref{fig1}), which has already been used to illustrate the 
performance of space-variant deconvolution methods (see, for 
instance, \citealt{denis,nagy}). 
The simulated image contains 470 stars on 
a range of 6 magnitudes with luminosity function and spatial distribution 
typical of a globular cluster; each of the stars has been convolved 
with a different PSF. A set 
of PSF images computed on a $5 \times 5$ grid is also included in the data set. 

We evaluated the goodness of our reconstruction by comparing the reconstructed 
image with the so-called ground truth, also available from the ftp, i.e. 
the true object that is the delta-function source model with no noise. 
Each source is represented by a pixel having a value equal to the source counts. 
The stars in the true object are positioned at integer pixel coordinates.

To distinguish between artifacts and stars, we built a threshold map computed 
by dividing the noise map of the simulated image by the maximum of the local normalized PSF. 
We obtained the noise map  by taking  both the photon noise due to 
the sources and the background and the Gaussian noise due to instrumental effects  into account
(i.e. RON). We estimated the noise map by means of the XNoise 
widget procedure included in the \texttt{StarFinder} program \citep{starfinder} 
for astronomical data reduction. The threshold map can  therefore be built by 
portioning the noise map using the same $5 \times 5$ PSF grid mentioned above and by 
dividing each obtained region by the maximum of the associated PSF.
This threshold map can be considered as a sort of noise map in the space of the reconstructed image 
(${\sigma}_{rec}$). It represents, point by point, the value of the reconstructed 
flux of the sources at the detection limit in the image space. 
The non-zero pixels in the reconstructed image having a 
value lower than this threshold map could have been easily generated by noise 
spikes. They are therefore classified as artifacts. Because of this criterion, 
all those stars in the reconstructed image that have a value lower than the threshold 
are non-detectable, even if they have counterparts in the true image. 
We considered three different 
thresholds (1, 2 and 3 times the noise map), and we analyzed three main 
quantities: 
\begin{itemize}
 \item{the number of lost stars defined as the number of pixels 
in the reconstructed image with a value smaller than the threshold and
corresponding to pixels of the true image with a value greater than 
the threshold; }
 \item{the number of false detections defined as the number of pixels 
in the reconstructed image with a value greater than the threshold and 
corresponding to pixels in the true image with a value equal to zero;}
 \item{the error in the reconstructed flux, expressed in magnitudes, 
evaluated comparing the true magnitudes of the input catalogue used to generate 
the true image with those derived from the counts in the corresponding 
pixels of the reconstructed image.}
\end{itemize} 
We report in Table \ref{table:1} the percentage of lost objects and false 
detections computed by considering the three different thresholds indicated above.
We computed the percentage of lost stars considering 
only the pixels of the true image having a value greater than the threshold.
In this case, where we performed no detection 
and there is severe crowding
in the GC centre, this method can cause an overestimation of the false detections.
In fact, the light spreading on the adjacent pixels due to a small shift in the centroid of the 
reconstructed object can originate a small group of false detections around the maxima, 
representing the true star.  To avoid this over-estimation, we should 
compare the relative maxima inside a certain aperture. This  would however cause
the blending of sources in the GC centre.   
We computed the percentage of  false detections  considering all the 
pixels of the reconstructed image having a value greater than the threshold. It 
is apparent that by fixing a reasonable threshold level established by the noise 
statistic of the image, the number of artifacts that could be confused for 
stellar sources is quite restrained.

\begin{table}
\caption{Detectability of the sources in the reconstructed image. The total number of sources in the true image is 470.} 
\label{table:1} 
\centering 
\begin{tabular}{c c c} 
\hline\hline 
Threshold & Lost & False\\ 
level & stars & detections\\
\hline 
1${\sigma}_{rec}$ & 2.5\% &  33.4\% \\ 
2${\sigma}_{rec}$ & 9.1\% & 2.6\% \\
3${\sigma}_{rec}$ & 20.6\% & $ < 1\%$ \\
\hline 
\end{tabular}
\end{table}

\begin{figure}
\centering 
\includegraphics[width=0.8\columnwidth]{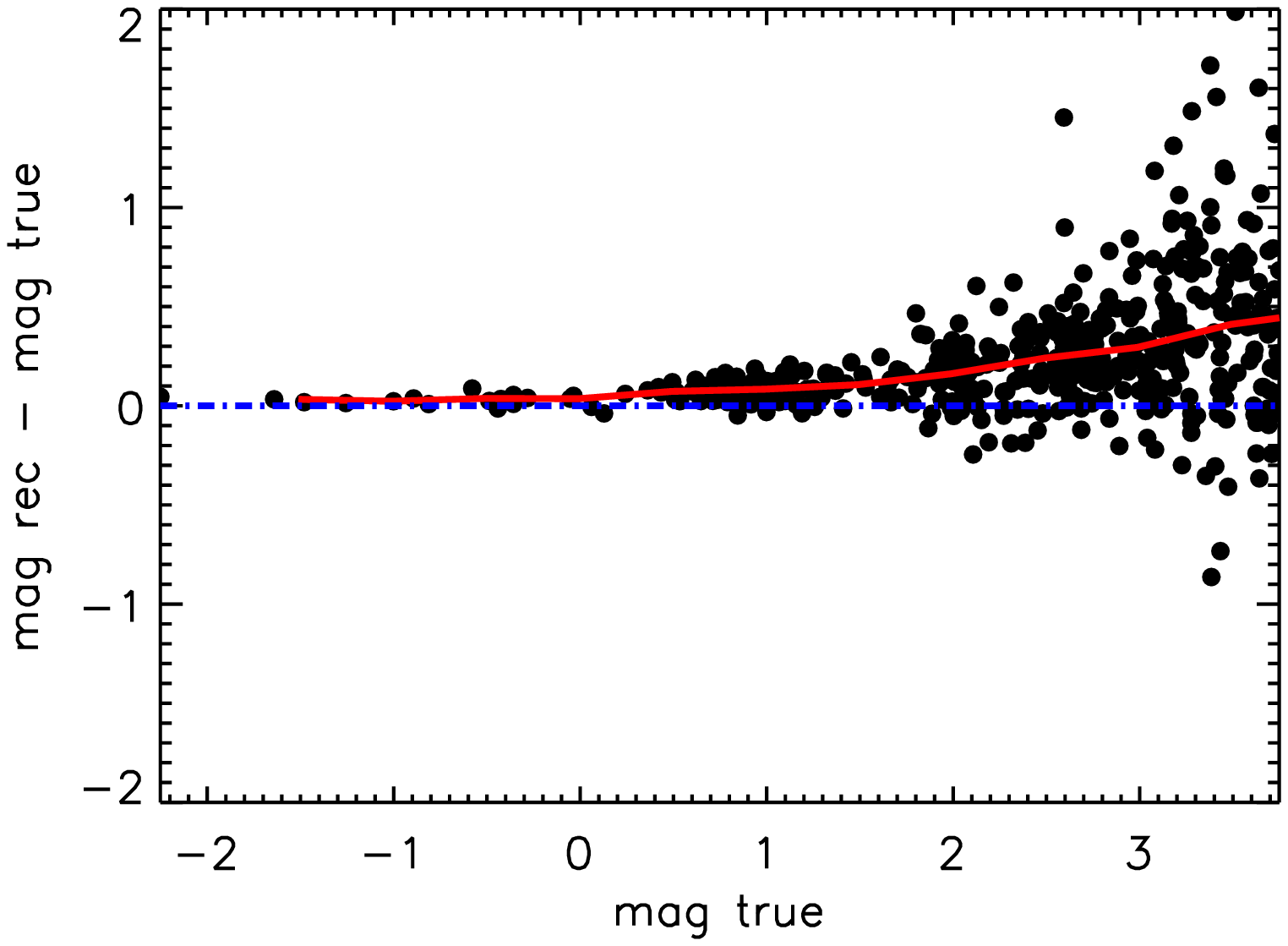}\\
\includegraphics[width=0.8\columnwidth]{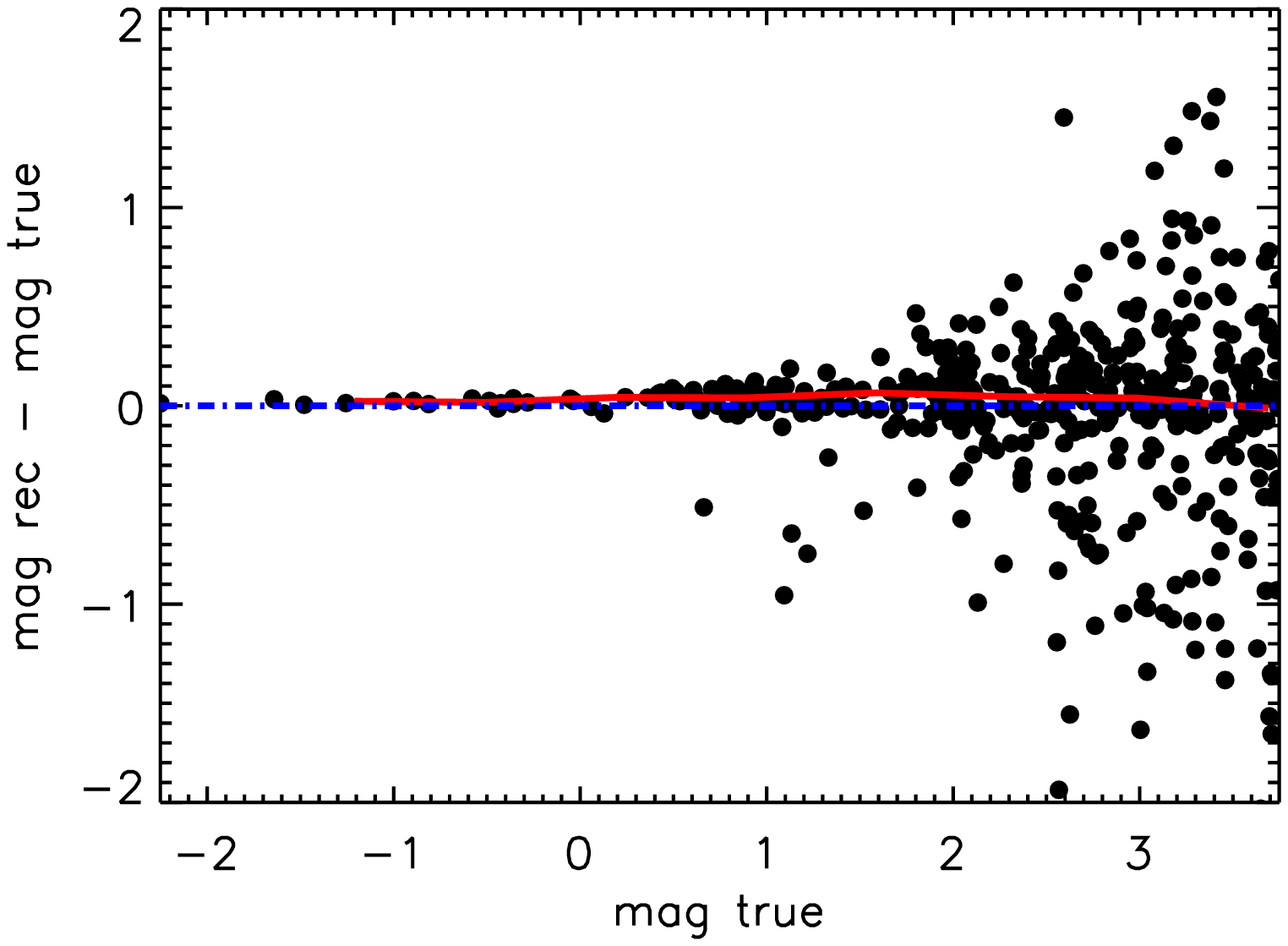}
\caption{Comparison of the true and reconstructed magnitudes. 
{\it Upper panel:} computed in correspondence of the true stars coordinate 
pixels. {\it Lower panel:} aperture photometry performed in a small region around 
the source coordinates. The red solid lines show the median error (bin amplitude = 0.5 mag).}
\label{fig1_2}
\end{figure}

The photometric error, due to the error in the source flux reconstruction, is 
illustrated in the upper panel of Fig.~\ref{fig1_2}. The error is computed comparing the counts of the 
true image and of the reconstructed image only in the input stars pixel 
coordinates. The counts that, for some reason, in the reconstructed image fall 
in the adjacent pixels, get lost. Because of this, the distribution of the magnitude 
differences shows an evident positively biased asymmetrical trend more prominent in 
the fainter region of the plot. 
This trend simply means that the reconstructed flux of the faint sources is spread on a small 
region of pixels rather than only one, consequently leading  to a small source 
location error (astrometric error). To compute more 
precisely the reconstructed flux, and so the photometric error, it is necessary 
to estimate it as the sum of the detected photons within an aperture of 
pre-fixed radius. This procedure has the drawback of not being  independent from 
the sources crowding, which is rather severe in the cluster core. 
The crowding causes the attribution of photons coming from stars closer 
than the aperture radius to the wrong source. Close stars are not clearly 
distinguishable and the signal belonging to many stars can be attributed to the brighter one, 
leading to a negatively biased asymmetrical distribution of 
the errors. This effect is called blending effect.
To reduce the blending effect, we adopted apertures with different sizes, depending 
on the distance from the GC core, and hence, on the crowding.
We used apertures from 7 down to 3 pixels in diameter in the external part of the GC, 
while we only considered  a one-pixel aperture in the cluster core. The median value of the 
photometric error depicted in the bottom panel of Fig.~\ref{fig1_2} (red line) is now 
close to zero along the entire magnitude range.


\begin{figure}
\centering
\includegraphics[width=0.85\columnwidth]{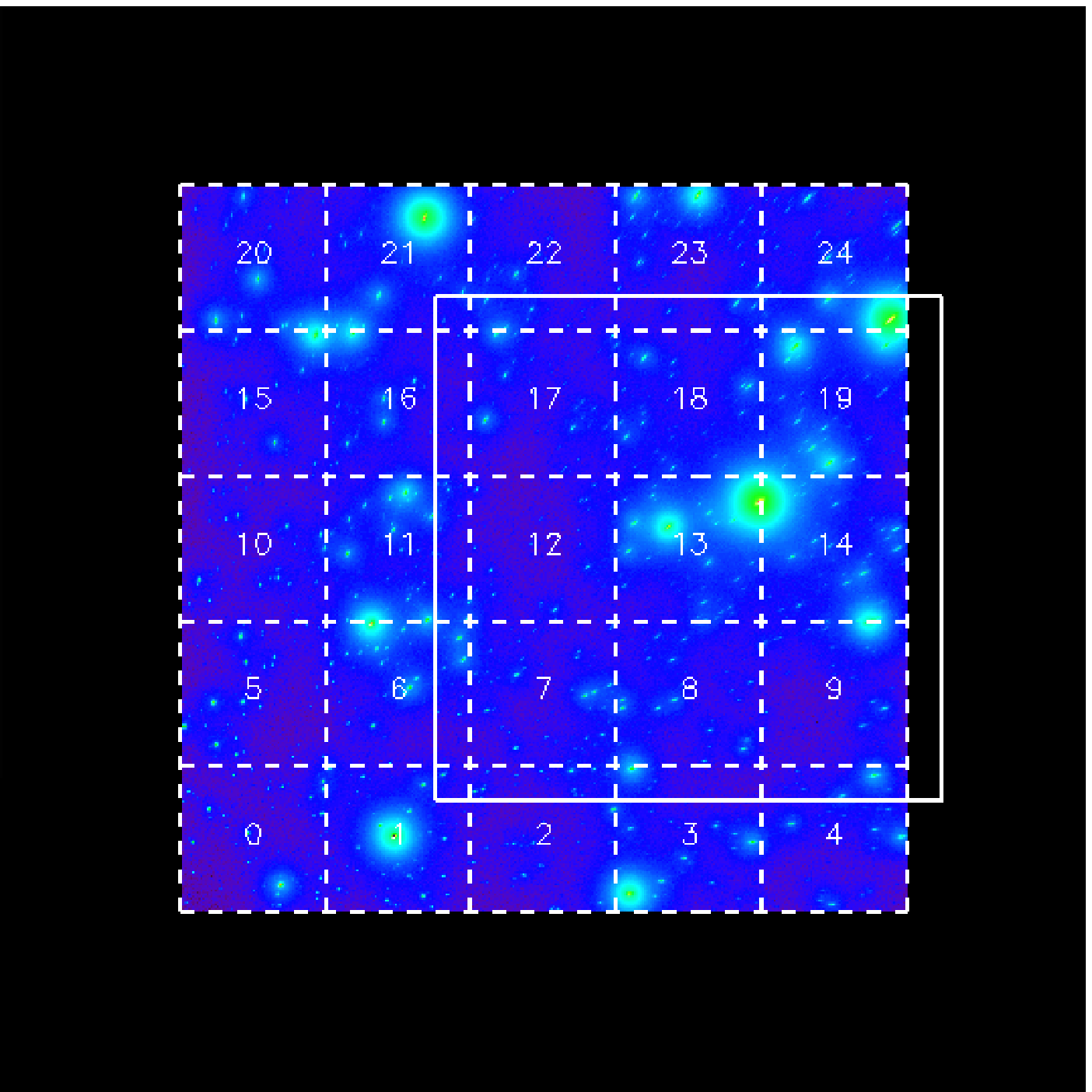}\\
\includegraphics[width=0.86\columnwidth]{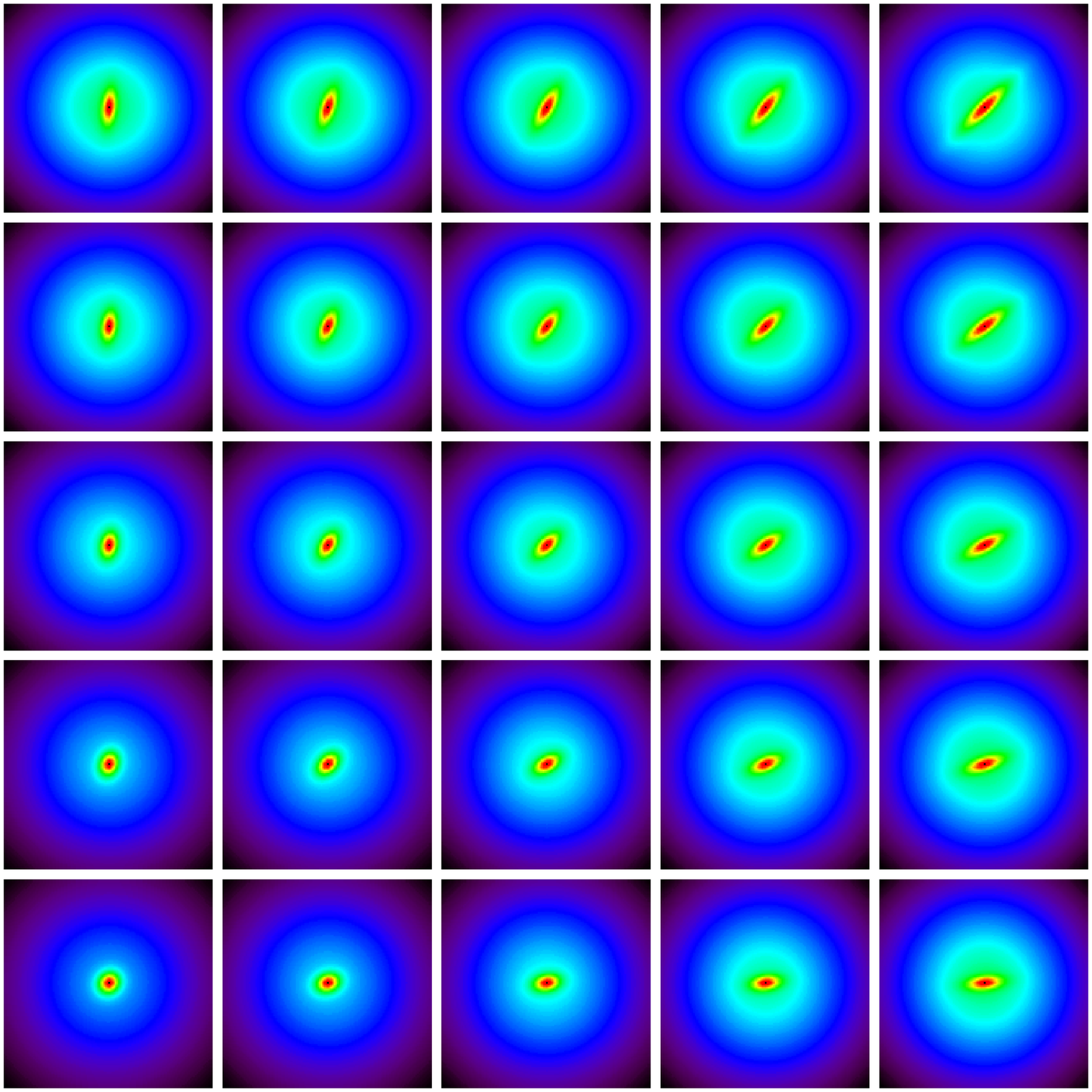}
\caption{Top panel: the sectioning of the input image (J band) domain into 5$\times$5 
overlapping regions; the domain nb 13 is overlaid  to show the 
overlap of the domains. Bottom panel: the 5$\times$5 grid of the PSF, computed 
from the model described in Appendix B.\label{fig3}}
\end{figure}

\section{Image simulation}
\label{aocase}

As in the previous case, images obtained with SCAO systems are characterized by 
structured PSF, with sharp core and extended halo, and by even more significant 
variations across the FoV. Up to now, none of the available codes for astronomical 
data reduction has been specifically designed to account for these PSF characteristics, 
which are typical of AO systems. 
The \texttt{StarFinder} code \citep{starfinder} was one of the first full
attempts to solve the problem of obtaining accurate photometry and astrometry from narrow field AO
images with highly structured, but spatially constant, PSF.
An effort in this direction has been reported in \citet{schreiber}, proposing an upgrade of 
the \texttt{StarFinder} code  to provide it with a set of tools to handle spatially 
variable PSFs. In the literature, other authors propose the space-variant 
deconvolution as a necessary tool for the exploitation of AO corrected images 
(i.e. \citealt{rusco, lauer}). In this panorama, AO images offer an interesting 
test bench for the proposed deconvolution method.

We therefore simulated the observations of an external region of a Galactic globular cluster (GC) 
in the $J$ band (central wavelength = 1.27 $\mu$m) and in the $K_s$ band (central wavelength = 
2.12 $\mu$m) with an 8 m class telescope equipped with a SCAO system. The images contain $\sim$ 
2800 sources on a range of about 10 magnitudes. The crowding ($\sim$ 6 stars per $arcsec^2$) of 
this field would be severe in seeing limited conditions, while it becomes moderate when using an
 AO system that shrinks part of the star light in a narrow diffraction limited core. However, the 
presence of the residual extended halo, which has a size comparable with the seeing, contributes 
to polluting the field image, enlarging the photometric error and making this case interesting to analyze. 
More details on the GC relevant parameters and on the characteristics of the image 
simulation are reported in Appendix A.

The adopted PSF has been modelled  to reproduce the main features of a typical 
SCAO residual PSF and its variation across the FoV. We considered a simple pure analytical 
model given by the combination of two 2D Moffat components: one representing the sharp 
diffraction limited core and the other the residual extended seeing halo. The adopted 
PSF model is described in detail in Appendix B.
An example of a simulated frame obtained by this PSF model is depicted in the top 
panel of Fig.~\ref{fig3}. 
The GS is situated just outside the FoV (bottom left corner). 
This choice maximizes the PSF variation across a moderate FoV. 
The Strehl ratio (SR) in both bands rapidly decreases across the FoV, ranging 
between $0.15$ and almost zero in the $J$ band and between $0.51$ and $0.18$ in 
the $K_s$ band. The maximum SR value corresponds to the GS position, while the 
minimum value occurs at the opposite image corner. This SR variation indicates 
a relatively small isoplanatic patch size compared to the FoV: $\theta_0 \sim 30 \arcsec$ in $K_s$ 
band and $\theta_0 \sim 15 \arcsec$ in $J$ band.

\section{Data reduction}
\label{reduction}

\subsection{Image deconvolution}
\label{img_dec}

Starting from the model described in Appendix B, we computed seven different sets of 
PSFs, namely 3$\times$3, 5$\times$5, 7$\times$7, 9$\times$9, 11$\times$11, 13$\times$13, 
and 15$\times$15. Each PSF has a fixed size of 512$\times$512 pixels ,  is positioned on a 
regular grid across the FoV of the image, and is centred on the centre of each corresponding sub-domain. 

In Fig.~\ref{fig3} we show both the input image (top panel) and the grid of the 
PSFs (bottom panel) for the 5 $\times$ 5 case. 
As mentioned in Sect.~\ref{sectioning}, the partial overlap of the domains has 
to be considered. In our simulations we adopt the 95\% of the EE of the PSF. 
In the case of $J$ band,  this corresponds to 510 pixels that is always the 
largest number with respect to all the differences $n_{p}-n$. In the other 
case ($K_{S}$ band), the extent of the PSF is 170 pixels, which is smaller or 
equal to the differences $n_{p}-n$. In Fig.~\ref{fig3} (top panel) a 
sub-domain is put in evidence to show this overlap in the case of J band.

We deconvolved the simulated images using ``Patch'' which, besides the input 
image, the number of sub-domains and the corresponding set of PSFs requires the 
background array (we assumed a constant background), 
RON and GAIN values (we assumed 20 e$^-$/pixel and unitary gain, 
see Table~\ref{table:A2}). We selected SGP pushed to convergence (iterations are 
stopped when the objective function described in Eq. \ref{KL} is approximately 
constant, according to a given tolerance, for instance 10$^{-9}$). Finally we set 
a maximum number of 5000 iterations to avoid a possible loop of the algorithm.

\begin{figure}
 \centering
\includegraphics[width=0.8\columnwidth]{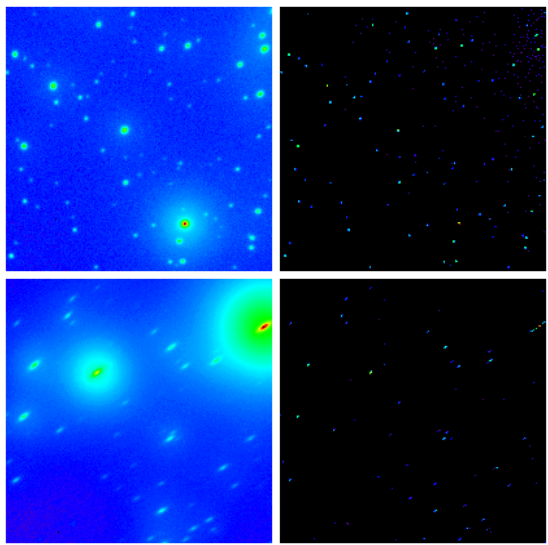}
\caption{Two sub-domains of the 5$\times$5 case before (left panels) and after 
(right panels) deconvolution. Referring to Fig.~\ref{fig3} for the sub-domains 
numeration, sub-domains number 0 and number 13 are represented 
in the top and bottom panels, respectively.}
\label{fig4}
\end{figure}

In Fig.~\ref{fig4} we show two sub-domains of the 5$\times$5 case before 
(left panels) and after (right panels) deconvolution. The sources in the 
reconstructed images look like delta-functions on a black background.  
The crowding effect due to the PSF extended halos has been largely reduced. 
Some of the reconstructed sources in the sub-domains more affected by the 
elongated PSF shape (and so far from the GS), show up a residual elongated 
pattern, especially close to the sub-domain edges. This is due to the mismatch 
between the local PSF (defined by an analytical continuous model that depends 
on the distance from the GS) and the PSF adopted for the sub-domain 
deconvolution (defined at the centre of the sub-domain). 
Artifacts, such as dotting, striping or ringing, are more concentrated along 
the sub-domains edges and around the brightest reconstructed sources. 
We recall that artifacts can be caused by different factors, like noise spikes 
or imperfect knowledge of the PSF. As a term of comparison, we also performed 
all the analysis by deconvolving the images with only one PSF (corresponding to
 the PSF at the centre of the images). We refer to this result as the 1$\times$1 
case. It represents our `reference case', where the proposed method of dividing 
the image in sub-domains and deconvolving each sub-domain with a local PSF is 
not applied. A crucial improvement of all the tested quantities is expected.  

\subsection{Stellar photometry}
\label{photometry}

Aperture photometry can be easily performed on the reconstructed images using 
apertures of few pixels. 
Unlike the procedure described in Sect.~\ref{hst_example}, where no detection has 
been performed in the reconstructed image, we assumed, as in the case of real data, 
that the positions of the stars in the field are not known.
Because of the absence of background (the median value in the reconstructed 
images is equal to zero) and  the delta-function shape of the reconstructed 
sources, the standard software packages commonly used to perform aperture photometry 
on astronomical images might be  inappropriate in this case. We therefore 
implemented an on-purpose package of IDL routines, which performs the source 
finding and the aperture photometry on the reconstructed image, returning a 
catalogue of fluxes and source coordinates. The objects are identified as relative 
maxima above a given threshold. The definition of the detection threshold in the 
space of the reconstructed image ($\sigma_{rec}$) is done as described in 
Sect.~\ref{hst_example}. We explored the amount of detected and lost objects 
with different confidence levels (1, 2 and 3$\sigma_{rec}$). 
We set the aperture diameters differently for each case, being the residual elongation of the sources 
dependent on the degree of discretization of the PSF. Also the number of generated 
artifacts around reconstructed sources is related to the local PSF estimation 
goodness, and hence, with the PSF discretization. Enlarging the aperture diameter allows 
us to include and recover the counts in the artifacts close to the reconstructed 
sources. The aperture diameters we adopted decrease from 13 pixels for the widest 
PSF grid step (3$\times$3 case) to 5 pixels for the finest grid steps (from 9$\times$9 
to 15$\times$15 cases). The flux of each source is computed by simply integrating 
the signal falling into the aperture centred in the source.  
The star positions have been computed as the centroid on a smaller aperture 
($3 \times 3$ for all the considered cases)  to avoid bias due to 
the presence of artifacts within the aperture. 


\section{Results}
\label{results}

\subsection{Reconstructed image analysis}
\label{analysis}
The reconstructed images are analyzed in terms of percentage of lost 
objects and number of artifacts.
 
To quantify the number of non-reconstructed objects,  
we performed aperture photometry in a very small 
region (3 pixels diameter) around the input source coordinates. 
If there are no photons within a given aperture centred in the coordinates 
of an object listed in the input catalogue (the one used to simulate the image), 
that object is classified as `lost', and so, not detectable.
Table \ref{table:2} collects the percentages of `disappeared' objects in the reconstructed 
image with respect to the total number of simulated stars. It is apparent that the 
number of detected objects becomes closer to 
the true number of objects by increasing the number of sub-domains, 
i.e. by reducing the difference between the actual PSF and the one used for the 
deconvolution of a certain sub-domain. It is interesting to note that if 
one consider more than 9$\times$9 sub-domains, the quality of the reconstructions 
does not improve. 

This is not a general result, but it depends on the adopted 
PSF and on its variation model across the FoV. Our model is characterized by a 
strongly varying core component, and also by a constant halo that contains a 
high percentage of the star signal, especially in the $J$ band where the SR is 
lower. The PSF variation, however, is higher in the $K_s$ band (see Figs. \ref{psfR} 
and \ref{psfF}). This choice allows us to test the boundary effects correction 
(thanks to the large, but constant halo), to verify the improvement of the image 
quality when approaching the actual PSF (thanks to the narrow and highly variable 
core), and to test the robustness of the algorithm to the PSF variation. It is 
also apparent that the $K_s$ band reconstructed image contains a lower percentage 
of lost objects, probably because of the higher signal-to-noise ratio (SNR), direct 
consequence of the higher SR.  

\begin{table}
\caption{Percentage of lost objects in the image reconstruction process in 
the $J$ and $K_s$ images.} 
\label{table:2} 
\centering 
\begin{tabular}{c c c} 
\hline\hline 
Number of& \multicolumn{2}{c}{Lost objects}\\
sub-domains & $J$ band & $K_s$ band\\ 
\hline 
$\bf{ 1 \times 1}$ & \textbf{40\% }&  \textbf{7\% }\\ 
$ 3 \times 3$  & 23\% &  3.5\% \\ 
$ 5 \times 5$  & 17\%  & 3\%  \\
$ 7 \times 7$ & 16\%  &  3\%  \\
$ 9 \times 9$ & 13\%  & 2.5\% \\
$11 \times 11$ & 13\%  & 2.5\% \\
$13 \times 13$ & 12\%  & 2.5\% \\
$15 \times 15$ & 12\%  & 2.5\% \\
\hline 
\end{tabular}
\end{table}

The overall performance seems to take advantage of a better PSF sampling 
across the FoV. This is also 
true in terms of photometric accuracy.  The number of artifacts that pollute 
the reconstructed image also decreases when the number of sub-domains increases. 
This is an expected result in terms 
of PSF mismatch, but not so obvious in terms of possible unwanted boundary 
effects that can rise when one splits the image into a large number of sub-images. 

To quantify the number of generated artifacts, a source detection has 
to be performed. The total number of detected objects is given by the sum of 
the real objects and the artifacts. Since both the reconstructed objects and 
the artifacts have a delta-function shape, we need to define a method to 
distinguish between them. As already mentioned, the real objects are identified 
as relative maxima above a given threshold. As a consequence, all those `objects' 
that have the maximum lower than the considered threshold are classified as 
`artifacts'. Thanks to this selection, we are able to recognize the majority 
of the spurious objects, but a residual number of artifacts remains hidden in 
the output catalogue, namely those artifacts that are above  
the threshold and have no candidate counterparts 
within 1 pixel distance in the input catalogue. 

Table~\ref{table:3} summarizes the percentages of false detections (or residual 
artifacts) with respect to the total number of detected stars in the two considered 
filters and for two different threshold values. It is encouraging to observe 
that the percentage of artifacts becomes very low for both bands ($\sim$2\% and 
$<$1\% in the $J$ and $K_s$ bands, respectively) when a 3$\sigma_{rec}$ threshold 
is adopted. This means that the majority of the spurious objects are relatively 
faint with respect to the real objects. 
When more than one image in the same 
filter is available, a match between different catalogues can filter out some of 
the remaining spurious objects, especially in low crowding conditions where the 
probability of ambiguous cases is low. 
Another interesting result is that the 
increasing sectioning of the image does not generate an enhancement of spurious 
sources. In the 15$\times$15 case, the sub-domain size is 69 pixels wide
and  the adopted PSF (512$\times$512 pixels) has a Moffat halo radius of 20 
pixels in the $J$ band (18 pixels in $K_s$) comparable to the sub-domain size.

\begin{table}
\caption{Percentage of false detected objects in the reconstructed $J$ and 
$K_s$ images. The percentage is computed with respect to the number of 
detected stars.} 
\label{table:3} 
\centering 
\begin{tabular}{c c c c c} 
\hline\hline 
Number of & \multicolumn{2}{c}{False detections}\\
sub-domains & $J$ band & $K_s$ band \\ 
 & $1\sigma_{rec}$ / $3\sigma_{rec}$ & $1\sigma_{rec}$ / $3\sigma_{rec}$\\ 

\hline 
$ \bf{1 \times 1}$ & \textbf{60 / 35 \%} & \textbf{26 / 13 \%} \\ 
$ 3 \times 3$  & 52 / 19 \% & 16 / 3 \% \\ 
$ 5 \times 5$  & 35 / 8 \% & 12 / 1.6 \%  \\
$ 7 \times 7$ & 34 / 7 \% & 11 / 0.8 \%  \\
$ 9 \times 9$ & 30 / 4 \% & 11 / 0.7 \% \\
$11 \times 11$ & 28 / 3 \% & 10 / 0.5 \% \\
$13 \times 13$ & 28 / 2 \% & 10 / 0.4 \% \\
$15 \times 15$ & 26 / 2 \% & 9 / 0.4 \% \\
\hline 
\end{tabular}
\end{table}

\begin{figure}
\centering
\includegraphics[width=0.94\columnwidth]{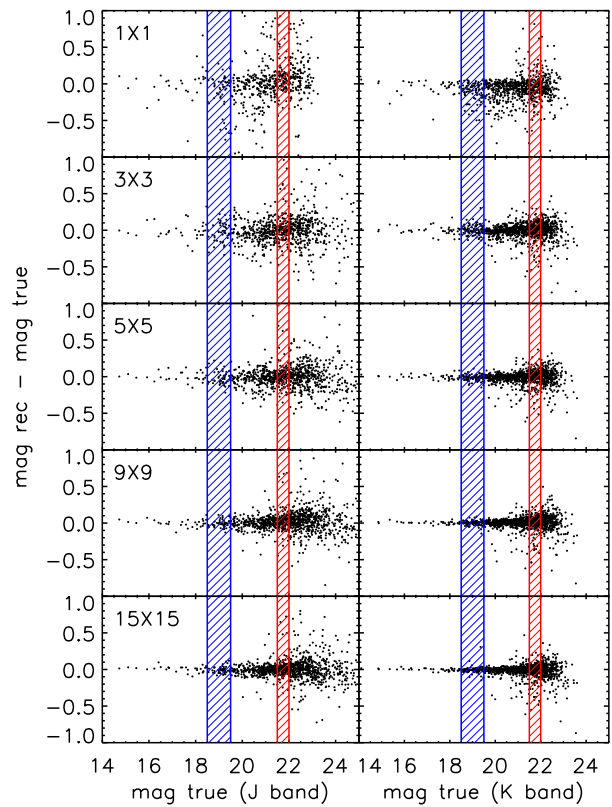}

\caption{Comparison of the input and measured magnitudes of the detected 
stars (photometric error) in the $J$ (left column) and $K_s$ (right column) 
bands for some of the analyzed cases. The detection threshold is set to 
3$\sigma_{rec}$. From the top: $1 \times 1$ (reference case), 
$3 \times 3$, $5 \times 5$, $9 \times 9$, and  
$15 \times 15$ sub-domains. The blue and red vertical stripes highlight two 
different bins of magnitudes with amplitude 1 mag (blue stripes) and 0.5 mag 
(red stripes). The photometric errors relative to the points falling in these 
stripes are reported in Table~\ref{table:4}.}
\label{fig5}
\end{figure}

\subsection{Photometric accuracy}
\label{accuracy}

The photometric accuracy is evaluated for each band and for each case by comparing the 
true magnitude of the detected stars with that measured from the reconstructed image. 
Fig.~\ref{fig5} shows a clear improvement in the reconstruction of the source fluxes 
if the number of sub-domains is increased so that the similarity between the actual 
PSF and the sub-domain central PSF is increased. This improvement seems to be crucial 
when one passes from the 3$\times$3 case to the 9$\times$9 case, where the gain starts 
to decrease considerably for both of the considered bands. The distribution of 
the magnitude differences looks nicely symmetrical for most of the considered magnitude 
intervals, and this means that the flux is preserved. As previously discussed, 
a flux loss would lead to a positively biased distribution. The reference 
case appears poorly populated as a consequence of the great number of lost objects.
We computed the photometric error, defined as the root mean square (r.m.s.) of the differences 
between the true and the estimated magnitude of the detected stars, 
in two different bins, one brighter and one fainter, and reported in Table~\ref{table:4}. 
The two considered bins are highlighted in Fig.~\ref{fig5} with the coloured vertical 
stripes. We chose a wider bin (1 magnitude) to compute the photometric error relative 
to the brighter sources  so that at least 50 sources would fall in the bin for 
each considered case. In the same two bins, we computed the astrometric error (see 
Table~\ref{table:5}), defined as the r.m.s. of the differences between input and 
recovered x-coordinate of the stars in a certain magnitude bin. Excluding the $1 \times 1$ 
case, which is reported as a term of comparison, there is no evidence of a strong 
dependence of the astrometric error on the number of sub-domains. To test for 
possible drawbacks due to the image sectioning and to our sub-domains approach, 
we also made the same analysis on a reconstructed image obtained by dividing the 
image in 15$\times$15 sub-domains and by deconvolving each sub-domain with the same PSF. 
For this purpose, we used the 1$\times$1 PSF, replicated one time for 
each sub-domain. The obtained photometric and astrometric errors are in a good 
agreement with the 1$\times$1 case for both  bands, as expected.  

\begin{figure*}
\begin{center}
\includegraphics[scale=0.85]{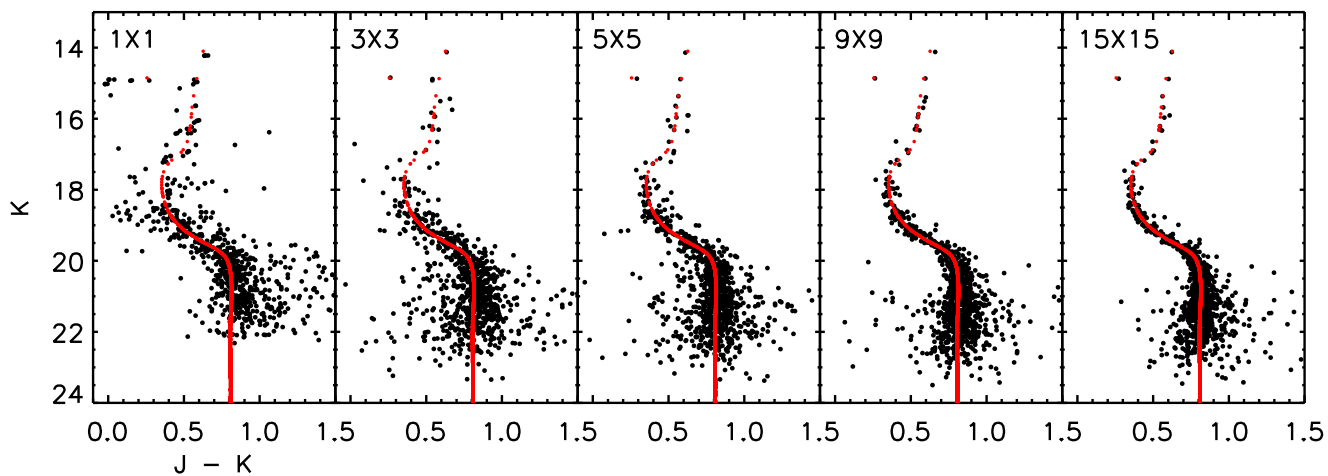}
\caption{Output ($K$, $J - K$) CMDs for the four image division in sub-domains cases depicted in Fig.~\ref{fig5}. The input CMD is over-plotted in red. The detection threshold is set to 3$\sigma_{rec}$ for all  cases.\label{fig6}}
\end{center}
\end{figure*}

\begin{table}
\caption{Photometric error in the $J$ (second and third columns) and $K_s$ (forth and fifth columns) bands computed considering two different bins (called `bin 1' and `bin 2' in the Table). The bins, highlighted in Fig.~\ref{fig5} in blue and red, are defined as follows: bin 1: 18.5 < mag true < 19.5 with amplitude = 1 mag (blue strip); bin 2: 21.5 < mag true < 22 with amplitude = 0.5 mag (red strip).} 
\label{table:4} 
\centering 
\begin{tabular}{c c c c c} 
\hline\hline 
Number of& \multicolumn{4}{c}{Photometric accuracy}\\
sub-domains & \multicolumn{2}{c}{$J$ band} & \multicolumn{2}{c}{$K_s$ band}\\ 
& bin 1 & bin 2 & bin1 & bin 2\\
\hline 
$ \bf{1 \times 1}$  & \textbf{0.17 }& \textbf{ 0.23} & \textbf{0.17} & \textbf{0.19} \\ 
$ 3 \times 3$  & 0.14 &  0.17 & 0.06 & 0.10 \\ 
$ 5 \times 5$  & 0.09 &  0.13 & 0.04 & 0.08 \\
$ 7 \times 7$  & 0.06 &  0.11 & 0.04 & 0.08 \\
$ 9 \times 9$  & 0.05 &  0.10 & 0.03 & 0.07 \\
$11 \times 11$ & 0.05 &  0.09 & 0.03 & 0.06 \\
$13 \times 13$ & 0.05 &  0.08 & 0.02 & 0.06 \\
$15 \times 15$ & 0.04 &  0.07 & 0.02 & 0.06 \\
\hline 
\end{tabular}
\end{table}

\begin{table}
\caption{Astrometric error in the $J$ (second and third columns) and $K_s$ (forth and fifth columns) bands computed by considering two different bins, as in Table~\ref{table:4}. The numbers are given in pixels.} 
\label{table:5} 
\centering 
\begin{tabular}{c c c c c} 
\hline\hline 
Number of& \multicolumn{4}{c}{Astrometric accuracy}\\
sub-domains & \multicolumn{2}{c}{$J$ band} & \multicolumn{2}{c}{$K_s$ band}\\ 
& bin 1 & bin 2 & bin1 & bin 2\\
\hline 
$\bf{ 1 \times 1}$ & \textbf{0.27} &  \textbf{0.28} & \textbf{0.07} & \textbf{0.15} \\ 
$ 3 \times 3$  & 0.05 &  0.14 & 0.02 & 0.12 \\ 
$ 5 \times 5$  & 0.05 &  0.13 & 0.01 & 0.11 \\
$ 7 \times 7$  & 0.05 &  0.12 & 0.01 & 0.11 \\
$ 9 \times 9$  & 0.03 &  0.10 & 0.01 & 0.10 \\
$11 \times 11$ & 0.03 &  0.11 & 0.01 & 0.10 \\
$13 \times 13$ & 0.03 &  0.10 & 0.01 & 0.10 \\
$15 \times 15$ & 0.02 &  0.12 & 0.01 & 0.10 \\
\hline 
\end{tabular}
\end{table}

To better evaluate the quality of the photometry, we combined the 
output catalogues in the $J$ and $K_s$ filters for each case.  Fig.~\ref{fig6} 
shows a selection of the obtained ($K$, $J - K$) colour-magnitude diagrams 
(CMDs) that refer to the same cases shown in Fig.~\ref{fig5}. The input CMD 
(see Fig.~\ref{fA1}) is over-plotted in red. The improvement of the 
photometric accuracy with increasing number of sub-domains is well represented 
by the gradual narrowing of all the cluster sequences in the CMD, going from the 
left to the right of the Figure. A slight quality improvement is also noteworthy 
between the last two depicted cases (9$\times$9 and 15$\times$15), mostly visible 
in the bright part of the CMD. This behaviour is confirmed by the photometric errors 
listed in Table~\ref{table:4}, where 
the values relative to the brighter bins decrease faster with respect to the 
sources in the fainter bins.
The bright sources are reconstructed better than the faint sources. 
The effect on the CMD depth is also apparent. The depth extends to fainter magnitudes, 
as a consequence of the decreasing number of lost objects, and therefore, of the increasing 
number of detected objects (see Table \ref{table:5}). The detection threshold for 
all the depicted CMDs is set to 1$\sigma_{rec}$. To appreciate the improvement of the 
overall quality of the CMDs, it is interesting to focus on some features of particular 
interest in scientific applications, like the Turn Off and the MS knee. The latter, 
in particular, is a powerful age diagnostic which has become accessible to observation 
only with the advent of modern AO systems \citep{bono}.
The CMD obtained by deconvolving the image with 15$\times$15 PSFs looks very narrow 
in correspondence of these two features, leading to a more precise fit of the observed isochrone. 


\section{Conclusions}
\label{conclusions}

We propose a sectioning method for reconstructing images 
corrupted by a space-variant PSF. First of all, the input image is sectioned in partially
overlapping sub-domains, the dimensions of which depend on the number, the extent, 
and the size of the PSFs. Then, each sub-domain (in which we assume that the PSF 
is space-invariant) is deconvolved with a suitable method with boundary effects correction. 

The effectiveness of the proposed method is proved by using two simulated 
images of stellar fields. The first image (Sect.~\ref{hst_example}) is an image 
of HST before COSTAR correction, well known in literature. We provide a deep 
analysis of the results we obtained, giving statistics and photometric error. 
The second test is the central part of this paper and is a simulation of a 
globular cluster in the $J$ and $K_{S}$ bands, characterized by a highly structured 
and variable PSF across the FoV (typical in AO). We  simulated 
the images by employing the continuous model, while we used several discrete 
grids of PSFs (from 1$\times$1 to 15$\times$15) in the data reduction process. 
In this sense, we also prove that the method is robust with respect to small 
variations of the PSF. 

In Sect.\ref{results}, we describe a method for distinguishing artifacts from 
reconstructed stars, since both have a delta-function shape. The number of 
artifacts can be controlled by a suitable threshold and, as shown in 
Table~\ref{table:3}, the number of false detections is very small ($\sim$2\% 
and $<$1\% in the $J$ and $K_s$ bands, respectively) if 3$\sigma_{rec}$ is chosen. 
Moreover, the number of artifacts and of lost objects decrease when the number 
of sub-domains increases, i.e. when the difference between the true PSF and the 
PSF used for deconvolving the sub-domain becomes smaller. We also report good 
photometric and astrometric results, again with increasing accuracy when the 
number of sub-domains increases. While there is no evidence of a dependence of 
the astrometric error on the number of sub-domains, the improvement in the 
reconstruction of the sources fluxes is crucial starting from the 
9$\times$9 case. Moreover, the CMD (obtained by the combination of the results 
in the two bands) is gradually narrowing and, especially in the 15$\times$15 case, 
the Turn Off and the MS knee are restored with excellent precision. All the relevant parameters used for the study of the 
reconstructed image quality (lost objects, artifacts and sources reconstruction) 
seem to agree on the optimal number of sub-domains to consider. 
This result, in terms of absolute number of sub-domains, depends mainly on the PSF 
variation amount across the FoV.  The adopted PSF is 
highly variable across the FoV. A softer PSF variation would lead to performance 
convergence with a smaller number of sub-domains.

A couple of remarks concludes our paper. The first  concerns the deconvolution 
methods performed in each section. In the Software Patch both RL and SGP algorithms 
are implemented. In our numerical tests we used SGP, which provides a speed up with 
respect to RL ranging from 10 to 20 and produces reconstructions with the same, 
sometimes better, accuracy. Since we considered point-like sources, we pushed the 
algorithm to convergence with a highly demanding stopping criterion and this 
demands time. For example the $J$-band 5$\times$5 case requires about 5.8 hours 
(a mean of about 14 minutes per sub-domain), using a personal computer with an 
INTEL~Core~i7-3770~CPU at 3.40GHz and 8~GB of RAM. Even if efficiency is not an 
issue we consider here, we indicate a few directions for reducing the 
computational time. First, the processing time can be certainly reduced by choosing 
a weaker stopping criterion. For example, in the mentioned case, by enlarging the 
tolerance from $10^{-9}$ to $10^{-5}$, the total processing time is reduced to 53 
minutes without changing  the quality of the reconstructed image too much: the 
number of detected objects slightly decreases, while the photometric error increases 
 about 2\%. Second, the sectioning method is quite naturally implementable on a 
multi-processor computer. Moreover, both RL and SGP implementation on GPU has already been 
considered \citep{prato2012}, showing that a speed-up of at least 10 is achievable 
with respect to the serial implementation. Therefore in 
the case of multi-processors and multi-GPUs a very significant speedup can be 
achieved, of the order of 100 with 10 GPUs.

The second remark is about PSF extraction and modelling. When handling real astronomical 
data, the local PSF is generally unknown. Two different approaches have been developed 
in  recent years: the PSF extraction and modelling from the data themselves 
\citep{schreiber,2013aoel.confE..78S}, and the PSF reconstruction technique 
\citep{1997JOSAA..14.3057V}. The first method involves only post-processing data 
operations, but it needs suitable stars well distributed across the FoV to model 
the PSF; the second one could imply the growth of the AO system complexity (especially 
when MCAO is involved), but it is independent of the observed field. Both methods are 
interesting when coupled with our deconvolution method, offering a robust and promising 
tool to reduce astronomical data characterized by a variable PSF. The application of our 
method to real data  using an estimated PSF is still under investigation and  will 
be published in a future paper.

The IDL code of the method described in this paper is available 
on the software section of the website 
\texttt{http://www.airyproject.eu}.

\begin{acknowledgements}
This work has been partially supported by INAF (National Institute for 
Astrophysics) under the project TECNO-INAF 2010 \emph{``Exploiting the adaptive 
power: a dedicated free software to optimize and maximize the scientific output
of images from present and future adaptive optics facilities''}. L.S. acknowledges 
Carmelo Arcidiacono and Antonio Sollima for the useful discussions.
\end{acknowledgements}

\bibliographystyle{aa} 
\bibliography{mybiblio} 

\begin{appendix}
\section{Globular cluster image parameters}
\label{appA}
\begin{figure}
\begin{center}
\includegraphics[scale=0.55]{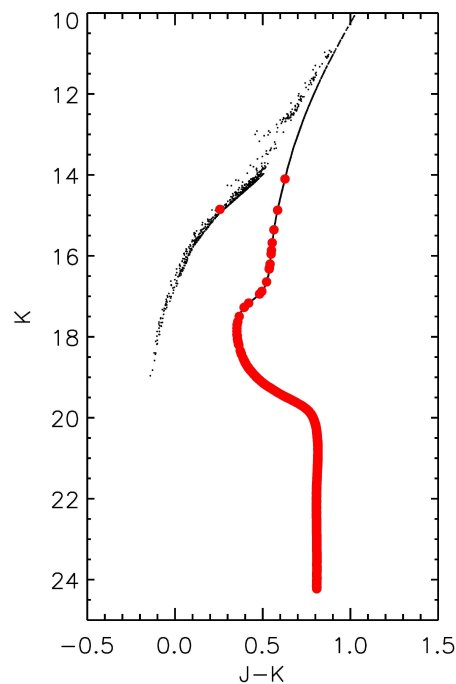}
\caption{The near-IR CMD of the GC stellar population. The total mass of the stars, 
formed 12 Gyr ago, is $5.4 \times 10^5$ M$_{\sun}$. The red dots highlight the 
sub-sample of the entire population that fall in the FoV considered in our simulation.\label{fA1}}
\end{center}
\end{figure}

The simulated GC is located at a distance of 10 kpc and is 12 giga-years 
old. We selected a region located at the half-mass radius of the cluster, at a 
distance of $128\arcsec = 6.2$ pc from the cluster centre. All the relevant parameters 
of the synthetic cluster are listed in Table \ref{table:A1}. 
The magnitudes and colours of the synthetic member stars have been drawn from theoretical 
models, the physical positions from N-body realizations of an equilibrium \cite{king1966} 
model. The observations were `acquired' with a fictitious, but realistic, 8.2 meters 
telescope equipped with a SCAO system + science camera + detector whose main characteristics 
are reported in Table \ref{table:A2}. These characteristics are very similar to the PISCES 
Infrared Imager with the Large Binocular Telescope Adaptive Optics System \citep{PISCES}.
A set of $J$ (central wavelength = 1.27 $\mu$m) and $K_s$ (central wavelength = 2.12 $\mu$m) 
band images have been simulated.  
The fraction of stars falling in the frames ($\sim 2800$ stars having $J \la 25$ mag 
and $K_s \la 24.2$ mag) is highlighted in red in Fig.~\ref{fA1}. To detect stars 
up to at least two magnitudes below the main-sequence (MS) knee, we fixed the total 
exposure time for each image to  1 hour for both bands. Therefore we computed the 
time for the individual exposures to avoid the saturation of any star. The $J$ band 
image is the result of the sum of 180 exposures of 20 seconds each, while the $K_s$ 
band image is the result of the sum of 240 exposures of 15 seconds each. 

\begin{table}
\caption{Parameters of the synthetic cluster. The assumed distribution is based on 
the King (1996) equilibrium model. HB and IMF mean  horizontal branch 
and initial mass function.} 
\label{table:A1} 
\centering 
\begin{tabular}{l c} 
\hline 
C (W0) & 1.9 (8.0) \\
Core radius  & 0.75 pc \\
Tidal radius  &  60 pc\\
N(stars) & $2 \times 10^6$ \\
 Stellar theoretical models & \citet{2008ApJS..178...89D} \\
Total Mass (stars)  & $5.4 \times 10^5$ M$_{\sun}$ \\
Age of Stars  &  12 Gyr\\                                       
 $[Fe/H]/[\alpha/Fe]$ & $-0.40 / 0.00 $\\
 IMF &  $N(m) \propto m^{-1.35}$\\
 HB: mean mass / $\sigma_{mass}$ & $0.60 / 0.04$ M$_{\sun}$ \\
 Binary Fraction & 0.0\% \\
 Assumed distance & 10.0 Kpc \\
 Assumed reddening  & E(B-V)=0.0 \\                                     
\hline 
\end{tabular}
\end{table}

\begin{table}
\caption{Telescope + Camera + Detector parameters adopted for the simulation. 
We assumed the average values of the sky surface brightness at Mount Graham 
\citep{2014NewA...28...63P}.} 
\label{table:A2} 
\centering 
\begin{tabular}{l c} 
\hline 
Collecting area &  50 m$^2$ \\
FoV & $21.5\arcsec \times 21.5\arcsec  $\\
Detector dimension  & $1024 \times 1024$ px \\
Pixel scale  &  $0.021 \arcsec$\\
Gain & 1 $e^{-}$/ADU\\
Read-out noise & 20 $e^{-}$ \\ 
QE  &  60\% in \textsl{J} band  \\
Saturation Level  &  $\ga$ 40000 ADU\\                                  
Dark current & $ 0.1$ $e^{-}$/sec  \\
Sky background mag &  15.82, 13.42\\
($J$ and $K_s$ bands) &  \\
\hline 
\end{tabular}
\end{table}

\section{PSF model}
\label{appB}
\begin{figure}
\centering
\includegraphics[width=0.87\columnwidth]{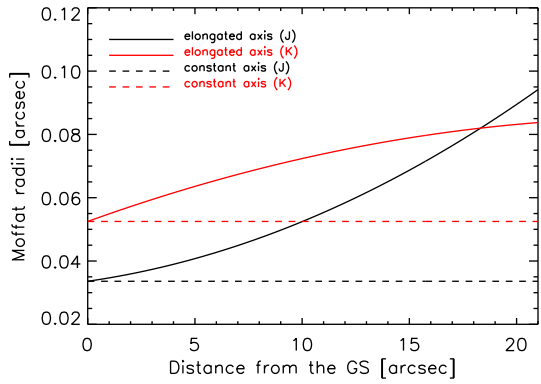}
\caption{Variation of the radii of the Moffat Core components of the PSFs in 
$J$ (black curves) and $K_s$ (red curves) bands across the FoV with respect 
to the GS distance in arcseconds. The continuous lines refer to the elongated 
radii of the Moffats. The dashed lines refer to the non-elongated radii of the 
Moffats. The non-elongated radii have been considered constant in this 
simplified model.\label{psfR}}
\end{figure}

\begin{figure}
\centering
\includegraphics[width=0.87\columnwidth]{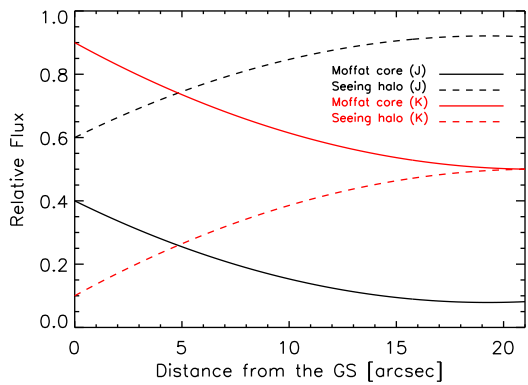}
\caption{Variation with respect to the GS distance of the relative flux contained 
in each of the two PSF components: core (continuous lines) + halo (dashed lines).\label{psfF}}
\end{figure}

At a first approximation, the AO PSFs can be approximated by the combination of 
different analytical components, such as Moffat, Lorentzian, or Gaussian 2D functions, 
their parameters varying with respect to the position in the FoV \citep{schreiber2012}. 
To simulate images with a continuous space-variant PSF, we
considered a simple pure analytical model given by the combination
of two 2D Moffat components:

\begin{itemize}
 \item Diffraction limited core: Moffat with a radial variation 
with respect to the guide star (GS) direction. The rotation angle reproduces 
the typical SCAO elongation pattern pointing towards the GS. The variation of 
the two Moffat half-light radii with respect to the distance from the GS is 
plotted in Fig.~\ref{psfR}. As depicted in Fig.~\ref{psfR}, the half-light 
radius pointing in the direction of the GS (i.e. along the elongated axis) is 
variable across the FoV and its variation is described by a polynomial function 
of the distance of the PSF location in the image from the reference position.
The half-light radius pointing orthogonally towards this direction has been considered 
constant and close to the diffraction limit.  
  \item Seeing halo: round Moffat (no elongation). The radius of this 
component (20 pixels in $J$ and 18 pixels in $K_s$) has been set  to 
reproduce a seeing disk of $\sim 0.6 \arcsec$ in $J$ band. 
 The only variable parameter in the FoV is the relative flux $F_{Halo} = 1 - F_{Core}$, 
where $F_{Core}$ is the relative flux contained in the core component. The halo 
component contains the signal due to the residual non-corrected atmospheric 
aberrations and, therefore, its relative flux grows with the GS distance following 
a second order polynomial trend.
Fig.~\ref{psfF} reports the variation of the flux distribution among the two PSF 
components (core and halo) in the two considered bands ($J$ and $K_s$) with 
respect to the GS distance.  
\end{itemize}   

\end{appendix}

\end{document}